\pdfoutput=1
\documentclass[11pt,twoside]{article}

\usepackage{asp2006}
\usepackage{epsf}
\usepackage{psfig}
\usepackage{lscape}
\usepackage{graphicx} %nrs inserted it.

\begin{document}
\title{What's So Peculiar About the Cycle 23/24 Solar Minimum?}   %%% Fill in title
\author{N. R. Sheeley, Jr.}   %%% Fill in author names
\affil{Space Science Division, Naval Research Laboratory, Washington DC 20375-5352, USA}    %%% Fill in author affiliations

\begin{abstract} %%% Abstract to run on from here.
Traditionally, solar physicists become anxious around solar minimum, as they await the high-latitude sunspot groups of the new cycle.  Now, we are in an extended sunspot minimum with conditions not seen in recent memory, and interest in the sunspot cycle has increased again.  In this paper, I will describe some of the characteristics of the current solar minimum, including its great depth, its extended duration, its weak polar magnetic fields, and its small amount of open flux.  Flux-transport simulations suggest that these characteristics are a consequence of temporal variations of the Sun's large-scale meridional circulation.
\end{abstract}

%%% MAIN BODY OF TEXT GOES HERE. CONSULT "INSTRUCTIONS FOR AUTHORS USING
%%% LATEX2E MARKUP", SECTIONS 2.3-2.6 FOR HELP WITH EQUATIONS, FIGURES,
%%% AND TABLES.

\section{Introduction}
When I was asked to give this talk, I wondered if this minimum was really peculiar,
or whether we were just feeling the anxiety that occurs toward the end of every
sunspot cycle as solar physicists await the first new-cycle active regions.

We are all familiar with this anxiety.  Flare researchers become anxious because they have contracts to study solar flares.  Energetic particle researchers become anxious because they want more data.  NASA managers become anxious because their spacecraft missions were justified in terms of what they would learn about solar activity.  But most anxious of all are the scientists who predicted the strength of the next sunspot cycle.  So when you encounter the forecasters, appreciate the stress they are under and be kind.

The level of anxiety increased during the 1976 minimum when Jack Eddy reminded us that sunspots became particularly scarce during the 70-year interval 1645-1715 and that another interruption of the sunspot cycle might occur at any time \citep{Eddy_76}.  His talks alarmed some people, and sent reporters to solar observatories to find out if we were headed into another Maunder Minimum.  We did not enter another Maunder Minimum in 1976 and we have emerged unscathed from two subsequent minima since that time.  Let's see how the present sunspot minimum compares with some of these earlier ones.

\section{The Sunspot Number}

Figure~1 shows the sunspot number during the interval 1895--2009.  The monthly means are plotted at the bottom of this panel with an arbitrary linear scale.  The natural logarithms of these monthly means are plotted at the top of the panel to show the lower numbers in more detail.  The logarithms are indicated by diamonds (or squares placed at -3 when the monthly means vanished, as happened for the most recent data point in August 2009).

\setcounter{figure}{0}
\begin{figure}[!h]
\centerline{\includegraphics[clip,scale=0.75]{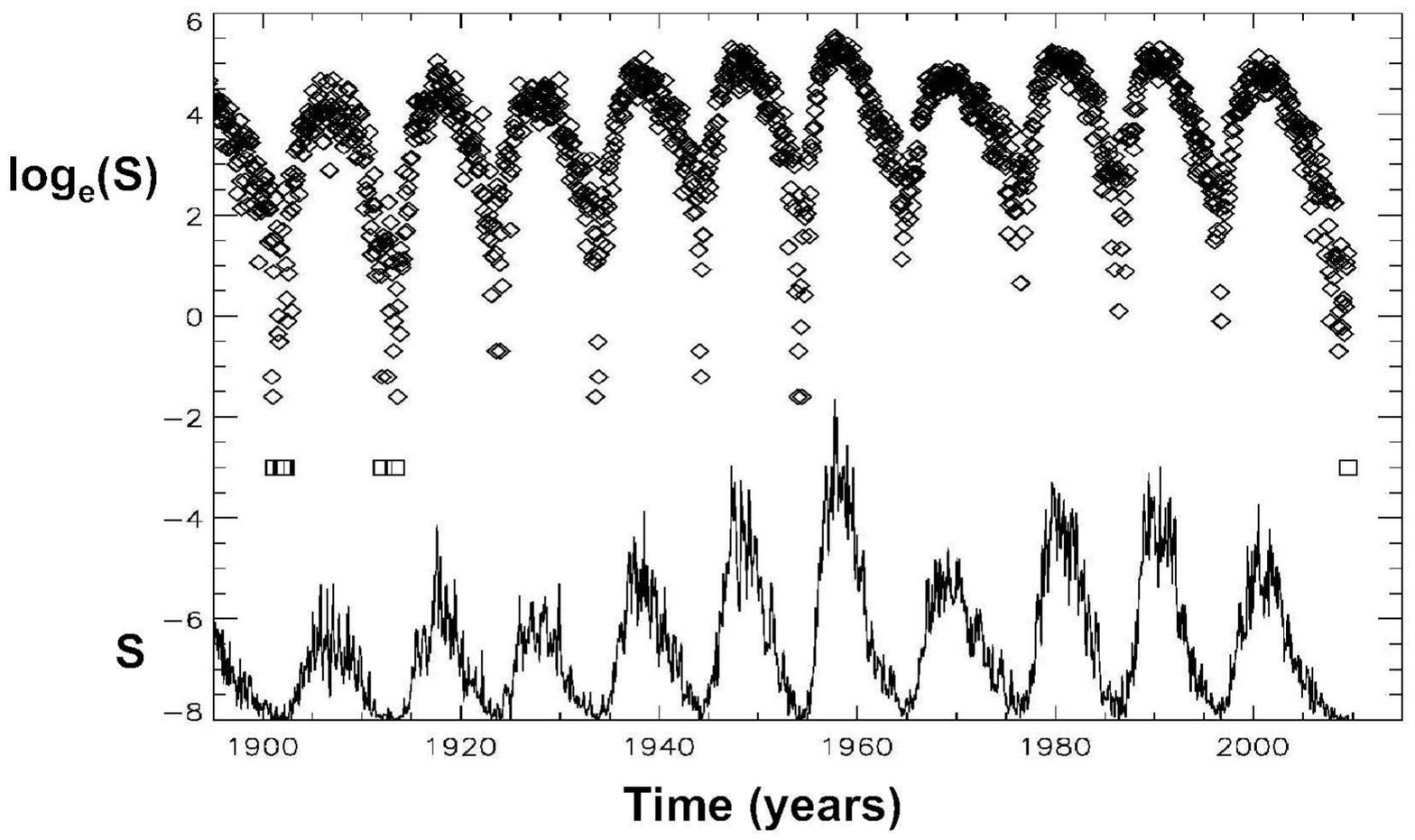}}
\caption{{\itshape Monthly averaged sunspot number (bottom) and its natural logarithm (top) plotted versus time in years, showing that the current minimum is comparable to those that occurred during the first half of the 20th century.}}
 \end{figure}

\begin{figure}[!h]
\centerline{\includegraphics[clip,scale=0.75]{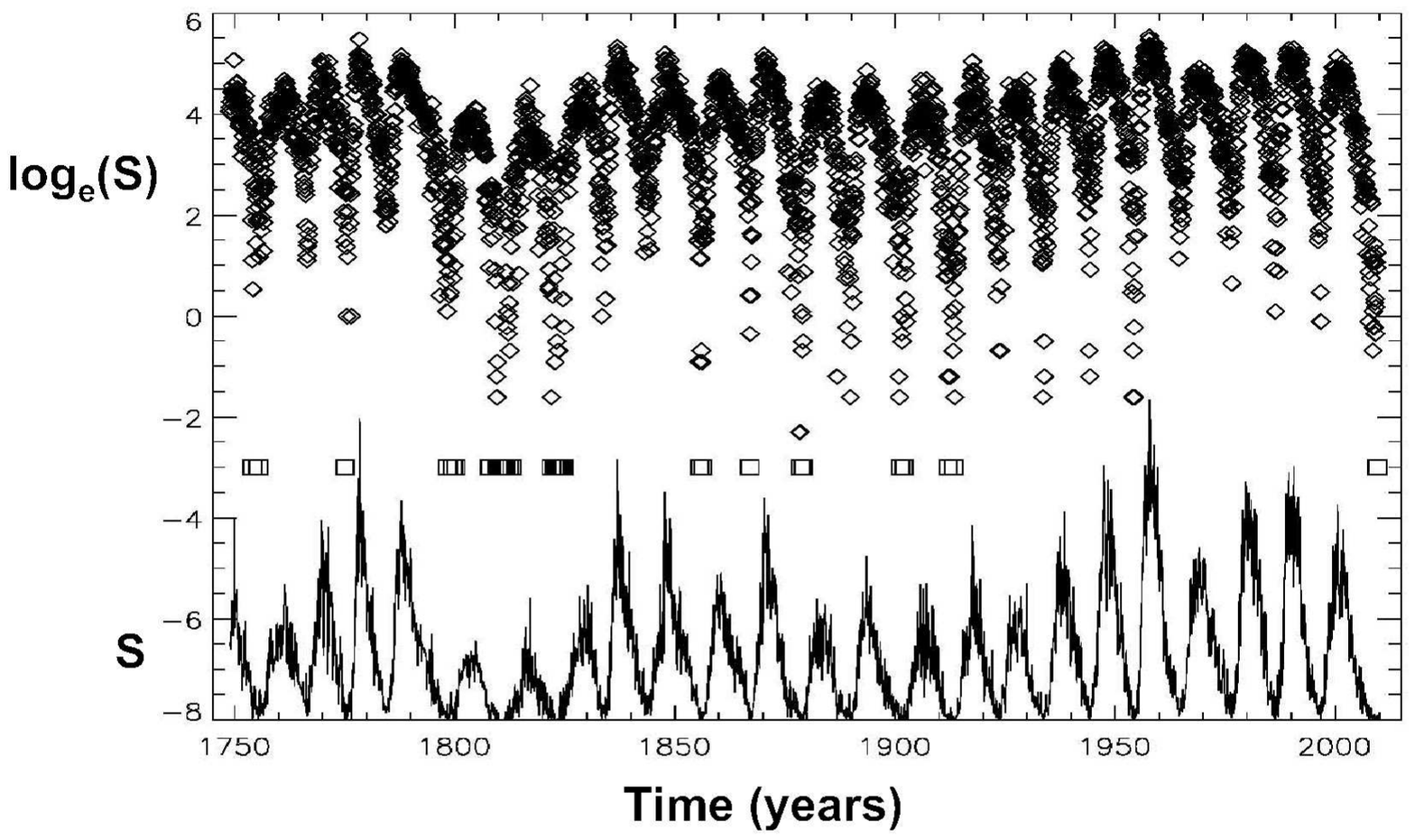}}
\caption{{\itshape Same as previous figure, except extending back to 1745.  Deep minima were common prior to the space age.}}
 \end{figure}

This figure shows that deep minima were common during the first half of the twentieth century, but that these deep minima suddenly disappeared after the very strong sunspot cycle in 1958.  Since that time the minima have been becoming progressively deeper.  The depth of the current cycle is now comparable to the depths of the cycles prior to 1958, and as low as the very deep minima in 1902 and 1913 if the August 2009 measurement is included.

Figure~2 provides a 250-year perspective, and shows that the series of deep minima extended back for 10 sunspot cycles before being interrupted by two shallow minima in 1833 and 1843.  Even deeper minima preceded them during the weak sunspot cycles in the Dalton minimum (1800-1830).  Thus, the present minimum is deeper than we have seen since the space age began, but not unusually deep on a time scale of 250 years or even 100 years.

\section{The Duration of the Minimum}

Figure~3 shows a time-latitude distribution of sunspot eruptions since 1875, prepared by David Hathaway(http://solarscience.msfc.nasa.gov/images/bfly.gif).  I have added the vertical dashed lines at the end of each cycle to help determine whether that cycle overlaps with the next one.

\begin{figure}[!h]
\centerline{\includegraphics[clip,scale=0.75]{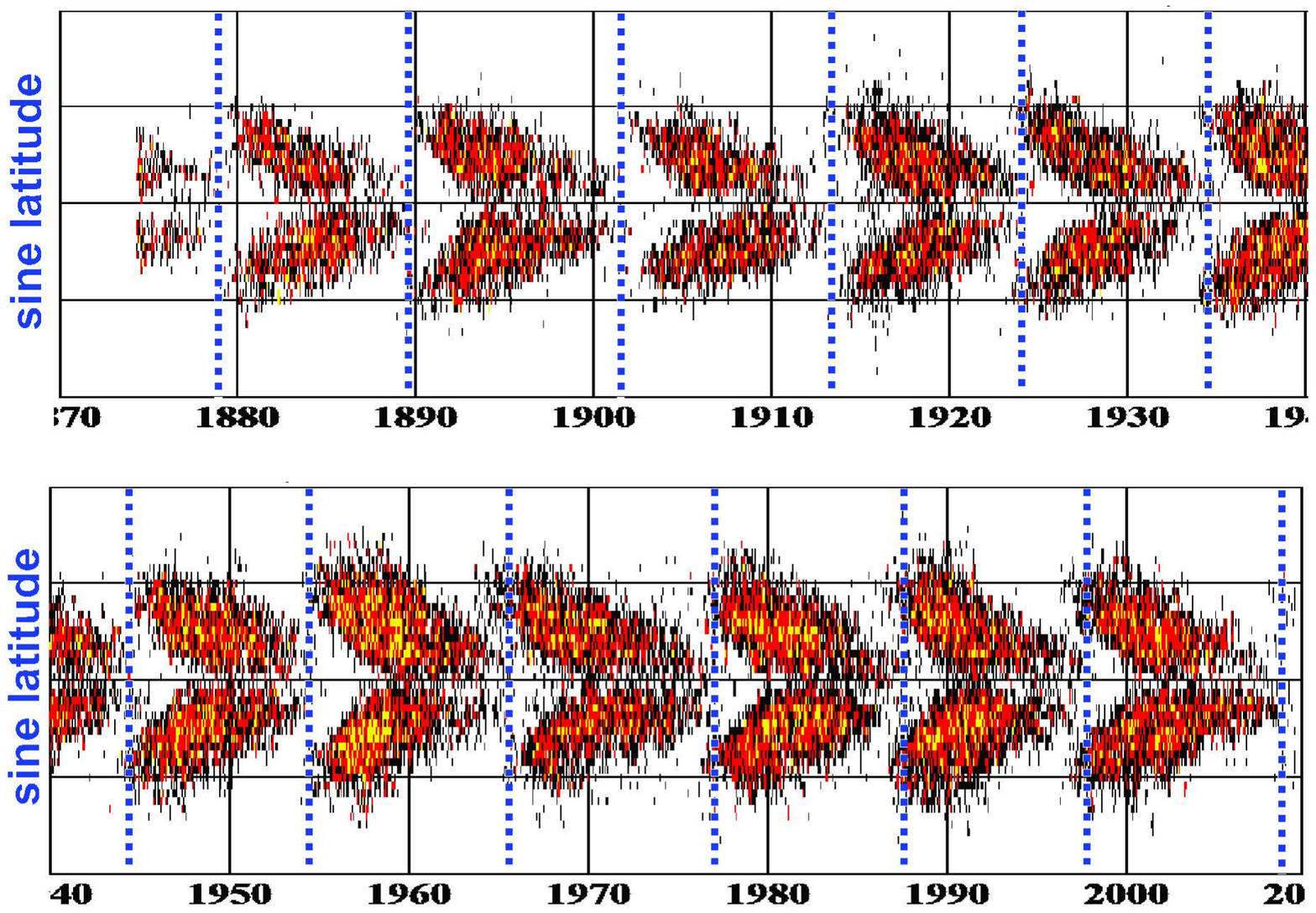}}
\caption{{\itshape The occurrence of sunspot groups as a function of time in years (horizontal axis) and sine latitude (vertical axis).  Vertical dashed lines are placed at the end of each sunspot cycle.  Since 1954, these lines have clipped the high-latitude wings of the next cycle, indicating no separation between consecutive cycles.  Exceptions include the present cycle and the southern hemisphere in 1966.}}
 \end{figure}

There is a clear, 1-year eruption-free interval at the end of cycle 23.  Moving backward in time, there are no comparable gaps between cycles until the minima in 1913 and 1902.  As indicated in Figure~1, These minima were especially deep, and they bounded the very weak sunspot cycle that peaked in 1906.  A shorter eruption-free interval occurred in 1954, separating cycles 18 and 19.

Examining the northern and southern hemispheres separately, we find that the southern hemisphere had an appreciable eruption-free interval in 1966, corresponding to a one-year delay in the start of sunspot cycle 20 in the southern hemisphere.

\begin{figure}[!h]
\centerline{\includegraphics[clip,scale=0.5]{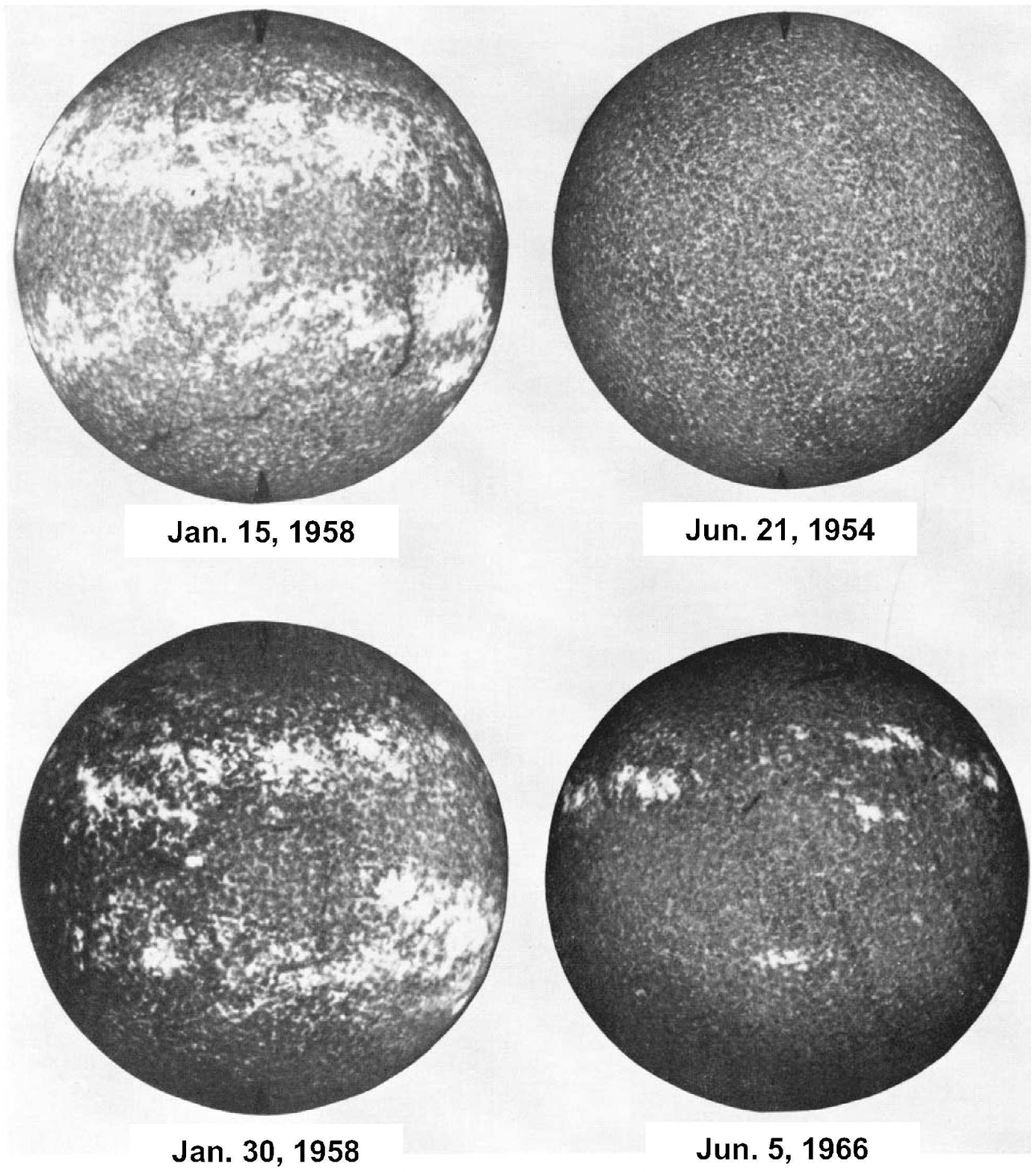}}
\caption{{\itshape Mount Wilson Observatory Ca II 3934 {\AA} images, showing the enhanced emission on the front (upper left) and back (lower left) sides of the Sun near sunspot maximum in 1958, the lack of such emission at sunspot minimum in 1954, and the north-south asymmetry at the start of the new cycle in 1966 \citep{SHE_67}.}}
 \end{figure}

This delayed start of southern-hemisphere activity is one that I experienced, but had forgotten until now.  The Ca II 3934 {\AA} spectroheliogram in the lower right panel of Figure~4 shows the north-south asymmetry on June 5, 1966 when I was on Kitt Peak, obtaining a time series of high-resolution K-line spectra of a region in the northern hemisphere.  During an exposure, the power to the telescope drive suddenly failed and the solar image drifted westward across the slit.  This produced an integrated K-line spectrum of the northern hemisphere.  When the power was restored, I made a similar scan across the southern hemisphere, this time intentionally.  I thought that the two spectra would be representative of conditions at sunspot maximum and minimum and therefore reveal all at once whether we could detect the sunspot cycle of the unresolved Sun from variations in its K-line emission, as Olin Wilson was starting to do for other stars \citep{WIL_78}.  

\begin{figure}[!h]
\centerline{\includegraphics[clip,scale=0.85]{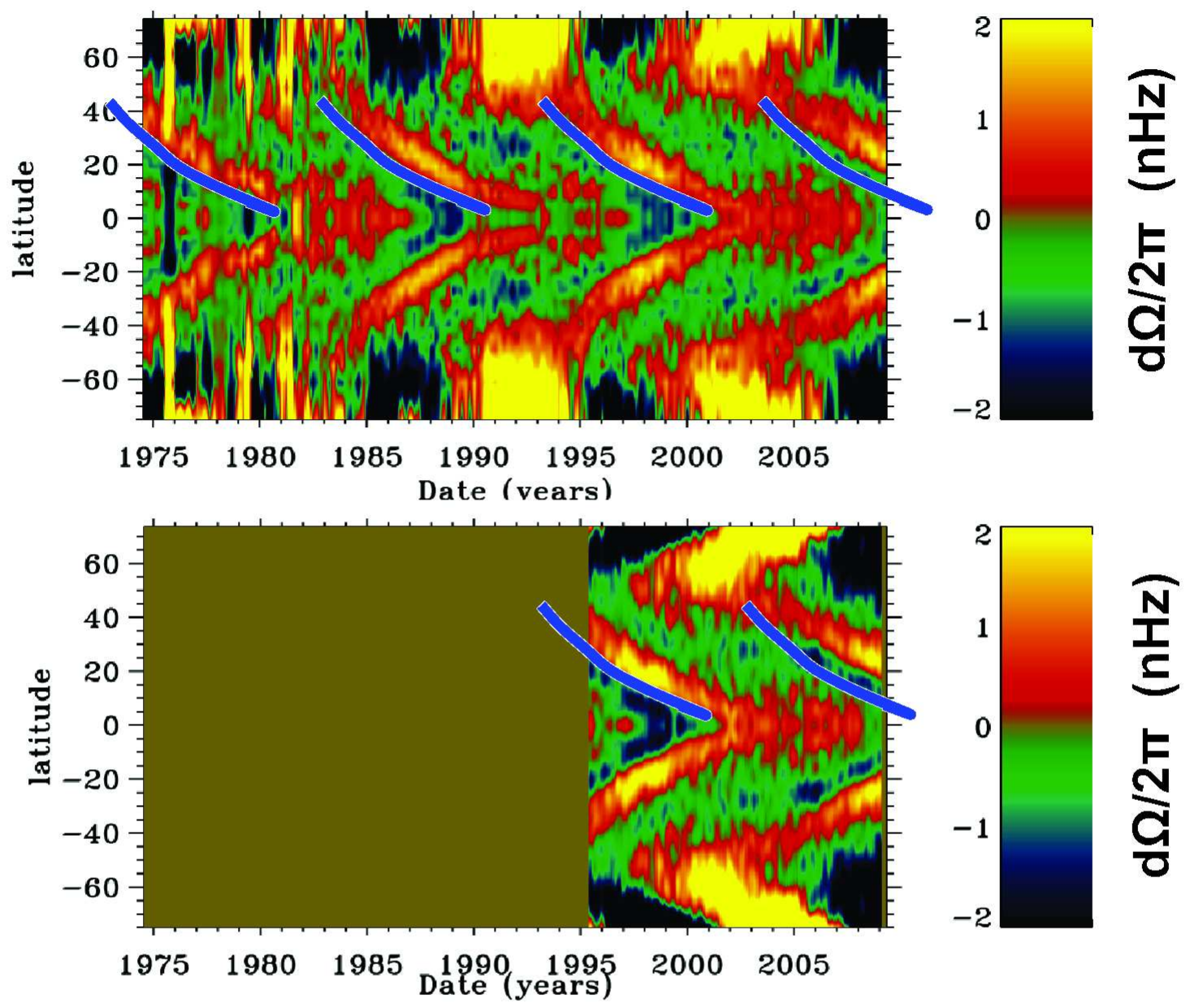}}
\caption{{\itshape Latitude-time displays of zonal flows obtained by subtracting the solar rotation profile from Doppler observations at the Mount Wilson Observatory (upperpanel) and from GONG/MDI global oscillation measurements (lower panel). Identical blue tracks fit the equatorial progressions during past sunspot cycles, but not the progression of cycle 24 \citep{HOW_09}.}}
\end{figure}

The so-called torsional oscillations \citep{HLAB_80} provide further evidence that the present sunspot minimum is different from the past three minima.  These features are alternating bands of prograde and retrograde rotation obtained when the long-term solar rotation profile is subtracted from the east-west component of the large-scale Doppler field (in the case of the Mount Wilson Observatory (MWO) measurements) or the global oscillation data (in the case of the Global Oscillation Network Group/Michelson Doppler Interferometer (GONG/MDI) helioseismic observations).  The lower branches of these residual east-west flows migrate toward the equator alongside (but not coincident with) the zones of sunspot eruption.

Figure~5 compares the MWO zonal flows observed during 1975--2009 \citep{UB_05} with the GONG/MDI flows \citep{HOW_09} during 1995--2009.  In this figure, a blue line was arbitrarily drawn along a boundary between the prograde and retrograde flows in cycle 23, and then shifted uniformly to the equatorial tracks in the other cycles.  As one can see, these identical blue tracks match all of the equatorward progressions except the one that began at high latitude in 2003.  The current migration has started more slowly than any equatorward migration since the observations began at Mount Wilson in 1975.  As \cite{HOW_09} first noted, this slow migration is a precursor to the delayed onset of sunspot cycle 24 and may provide another clue to the origin of the delay.    

\section{Weaker Polar Magnetic Fields}

We have all heard that the polar field is weaker now than it has been for many years.  Figure~6 shows the Wilcox Solar Observatory (WSO) measurements since 1976.  The polar field is about two-thirds as strong as it was during the previous minimum.  The axial component of the Sun's dipole field shows a corresponding decrease, suggesting that the open flux is also about two-thirds of its previous value.

\begin{figure}[!h]
\centerline{\includegraphics[clip,scale=0.85]{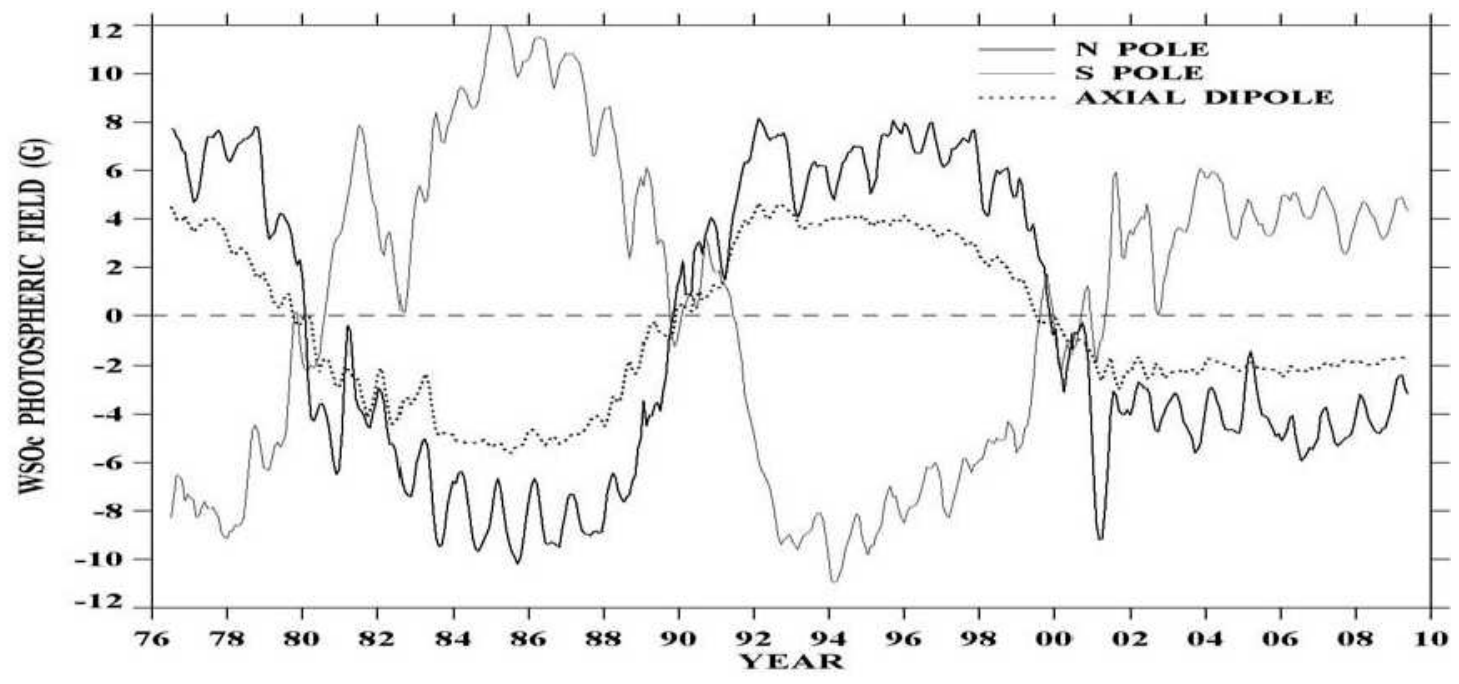}}
\caption{{\itshape Plots of the Sun's polar magnetic fields and axial dipole derived from observations at the Wilcox Solar Observatory during 1976--2009.}}
 \end{figure}

\begin{figure}[!h]
\centerline{\includegraphics[clip,scale=0.8]{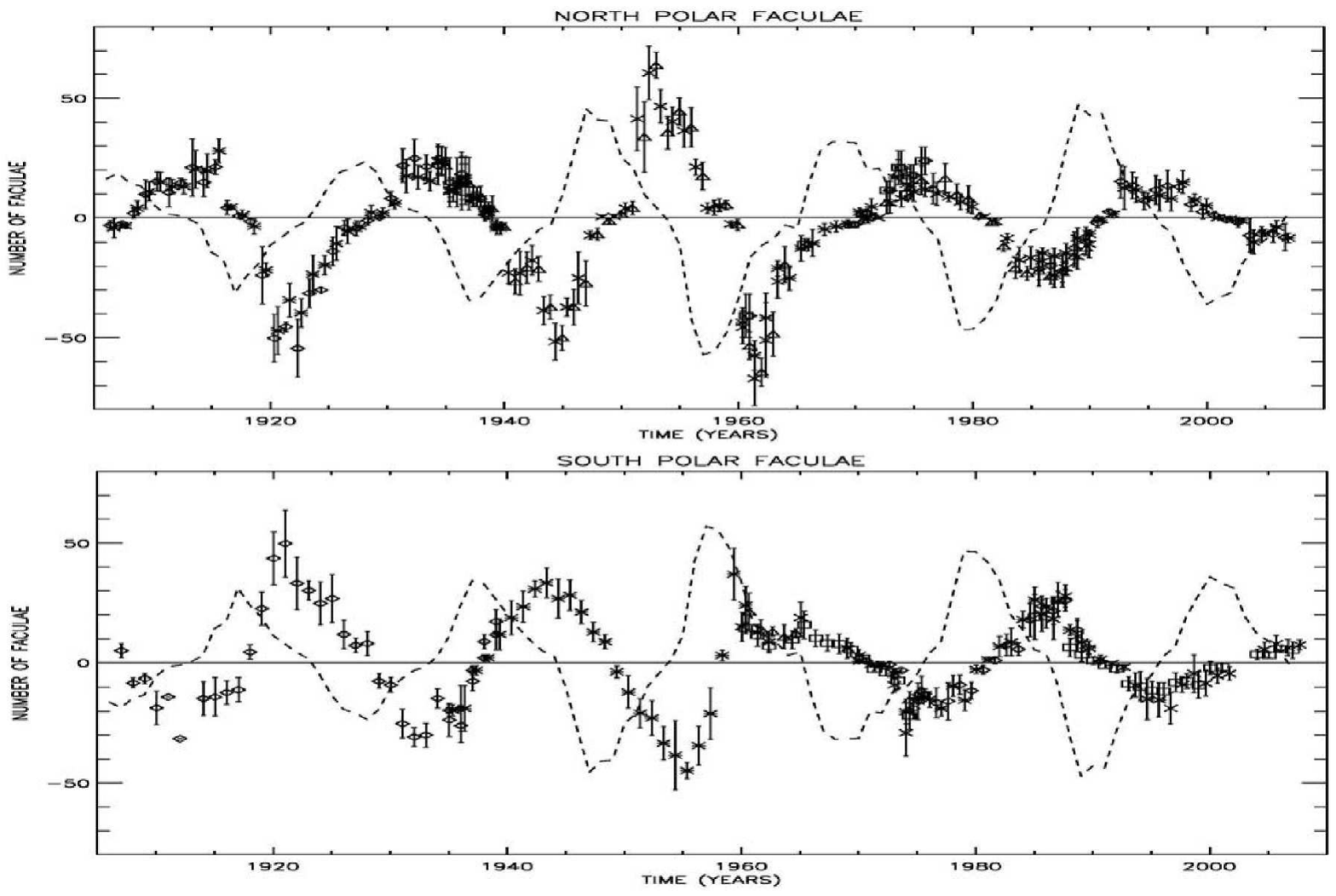}}
\caption{{\itshape The numbers of north and south polar faculae during their times of greatest visibility (fall or spring), and the yearly sunspot number for the full disk, multiplied by 0.3, and assigned the polarity of the following spots in each hemisphere (dashed lines) \citep{SHE_08}.}}
 \end{figure}

It is interesting to see how far back in time we can extend these polar field measurements.  Polar faculae are visible on white-light images obtained daily at the Mount Wilson Observatory since 1906, and their numbers provide a reliable indication of the polar magnetic field strengths.  Figure~7 shows the numbers of polar faculae counted during the favorable intervals of each year (fall or spring) since 1906.  These numbers have been assigned the polarities of the corresponding polar magnetic fields (since the invention of the magnetograph in 1952), or extrapolated smoothly through zero (in the premagnetograph years).

This figure suggests that the polar fields are weaker now than they have been in the last 100 years.  Rapid changes in the early 1960s are probably due to alternating bands of flux carried poleward by meridional flow.  Aside from these changes, this cycle of southern-hemisphere faculae was the second smallest in this 100-yr record.  We have already seen in Figures~3 and 4 that the southern-hemisphere activity was delayed by about one year in 1966.  Is it a coincidence that the delayed onset of this activity and the delayed onset of cycle 24 were both preceded by unusually weak polar fields?  This provides a motive for reexamining the white-light images during the extended minimum around 1913 to see how weak the polar fields may have been during that time.

\section{Less Open Magnetic Flux}

During the declining phase of the sunspot cycle, the eruption of flux creates and
maintains low-latitude coronal holes with accompanying warps of the streamer belt.  These coronal holes gradually die out at sunspot minimum, as the old-cycle eruptions stop and the relatively strong polar fields grab the dwindling remnants of open field lines at low latitude.  During the present minimum, the old-cycle eruptions stopped early in 2008, but the low-latitude holes and the warped streamer belt persisted for at least another year.  This peculiarity is due to the relatively weak polar magnetic fields \citep{WRS_09}, as one can see from the experiment performed in Figure~8.  

The left panels show the NSO photospheric field (top), the observed Fe XII 195 {\AA} emission (second from top), the photospheric distribution of open flux derived from a potential field extension of the observed field, and the corresponding derived field at 2.5$R_\odot$.  The map of open flux shows colored areas at low latitude that correspond to dark coronal holes in the map of Fe XII 195 {\AA} emission.  The right panels show the same maps with polar fields that are twice as strong (12 Gauss compared to 6 Gauss).  This change caused the derived regions of open flux to disappear from low latitude and the neutral line of the coronal field to flatten toward the equator.  The same experiment showed very little change nine rotations earlier when large active regions were still present at low latitude.

The survival of a low-latitude coronal hole depends on its field strength relative to the field strength of the polar hole of opposite polarity.  Consequently, the low-latitude holes last longer when the polar fields are weak.  Also, if the polar field strengths differ appreciably in the two hemispheres, then the longer-lived low-latitude holes would have the polarity of the stronger polar field.  This allows us to predict that the surviving low-latitude holes ought to have had negative polarity in 1966 when the weaker south polar field had positive polarity.
 
\begin{figure}[!h]
\centerline{\includegraphics[clip,scale=0.7]{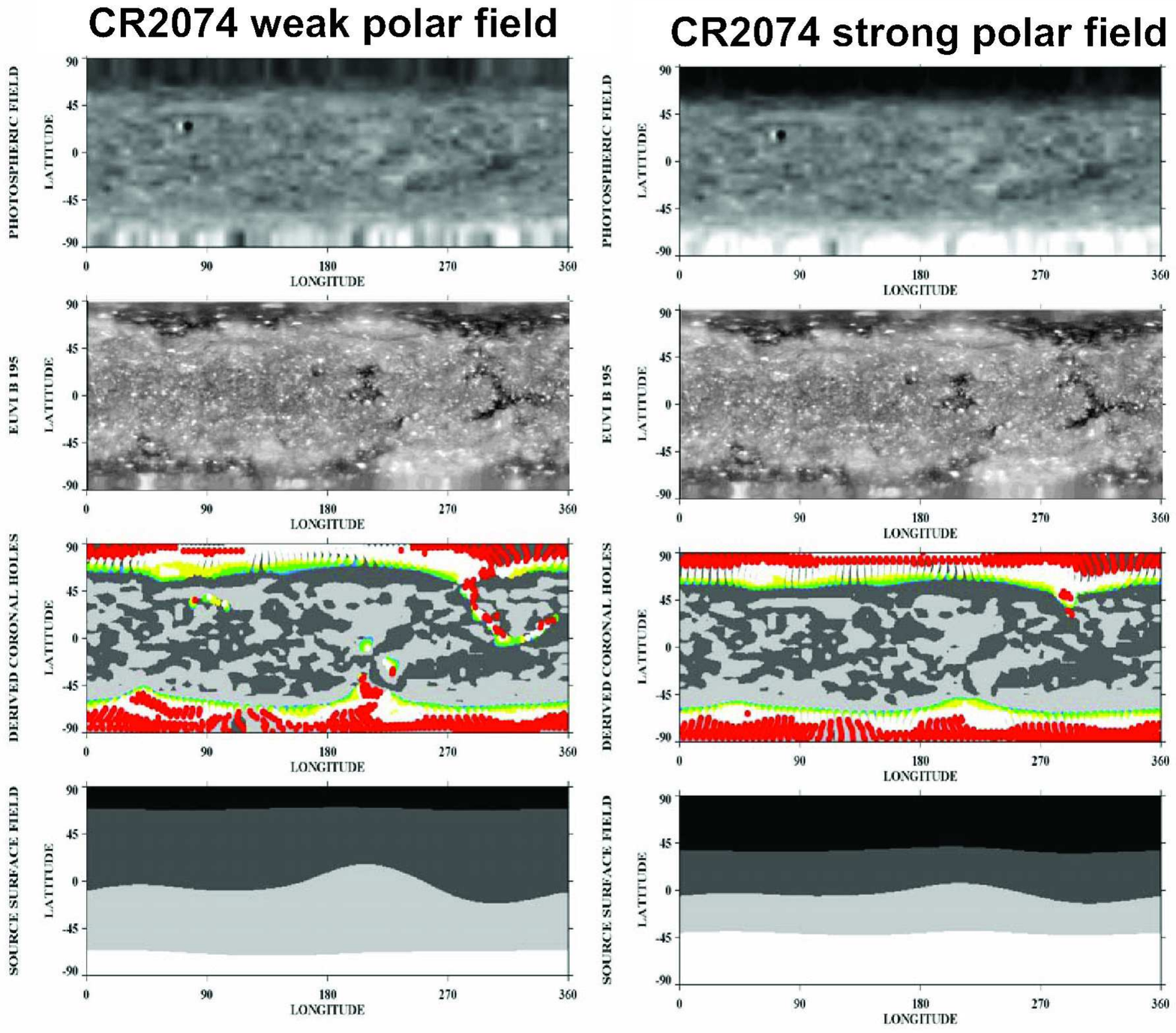}}
\caption{{\itshape Carrington maps of photospheric field (top), Fe XII 195 {\AA} intensity, derived open flux, and source-surface field (bottom), showing that the low-latitude coronal holes disappear and the source-surface neutral line flattens when the polar field strength is increased from 6 G (left) to 12 G (right).}}
 \end{figure}

Figure~9 compares the total open flux on the Sun, derived from potential field extrapolations of the observed photospheric magnetic field, with the total open flux, derived from in situ measurements of the radial component of the interplanetary magnetic field.  These overlapping curves show similar behaviors with low values during each sunspot minimum, when nearly all of the flux originates in the polar coronal holes, and high values near and after sunspot maximum.  These high values occur when flux erupts in longitudinal phase and increases the strength of the equatorial dipole.  The individual peaks decay with a lifetime of about 1.5 years as meridional flow carries the flux to midlatitudes where it is sheared by differential rotation and dissipated by supergranular diffusion. As indicated by the arrow, the present amount of open flux on the Sun and in the heliosphere is the lowest since observations began in 1967.

\begin{figure}[!h]
\centerline{\includegraphics[clip,scale=0.7]{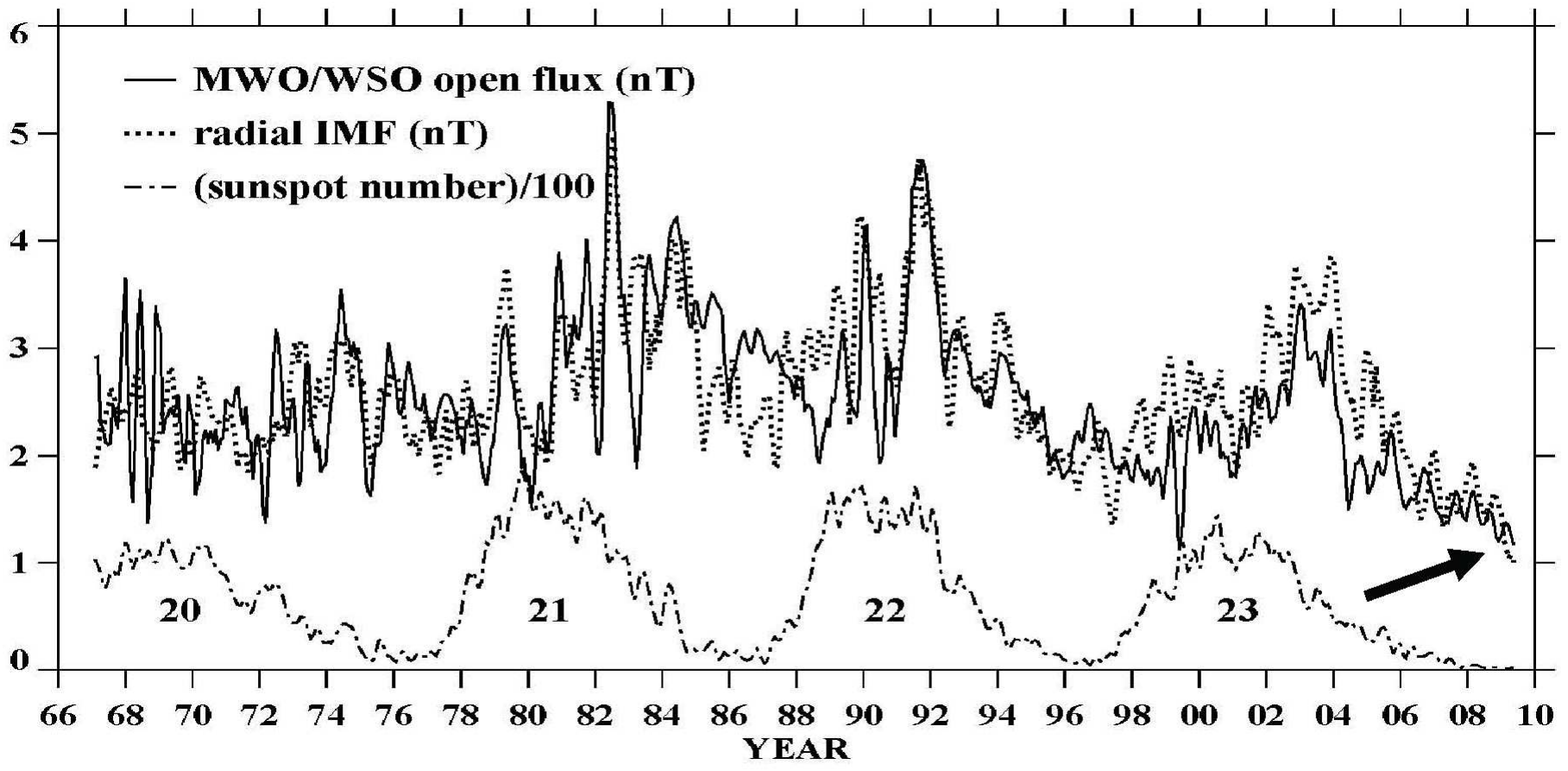}}
\caption{{\itshape The total open flux on the Sun derived from photospheric magnetograms (solid line) and derived from in situ measurements of the radial magnetic field (dashed line).  For comparison, the sunspot number is plotted below (dashed-dot line). The current value (arrow) is the lowest since observations began in 1967.}}
 \end{figure}

\section{The Effects of Meridional Circulation}

Figure~10 shows a longitudinally averaged map of photospheric magnetic field measurements obtained at the Mount Wilson Observatory (MWO) since 1967.  From the first 13 years of these observations, \cite{HLAB_81} identified `episodic poleward surges' of flux extending from the sunspot belts to the poles of the Sun.  They argued that these surges were evidence for a poleward meridional flow because supergranular diffusion by itself \citep{RBL_64} would blur out the flux distribution and not show these concentrated streams.

\begin{figure}[!h]
\centerline{\includegraphics[clip,scale=0.7]{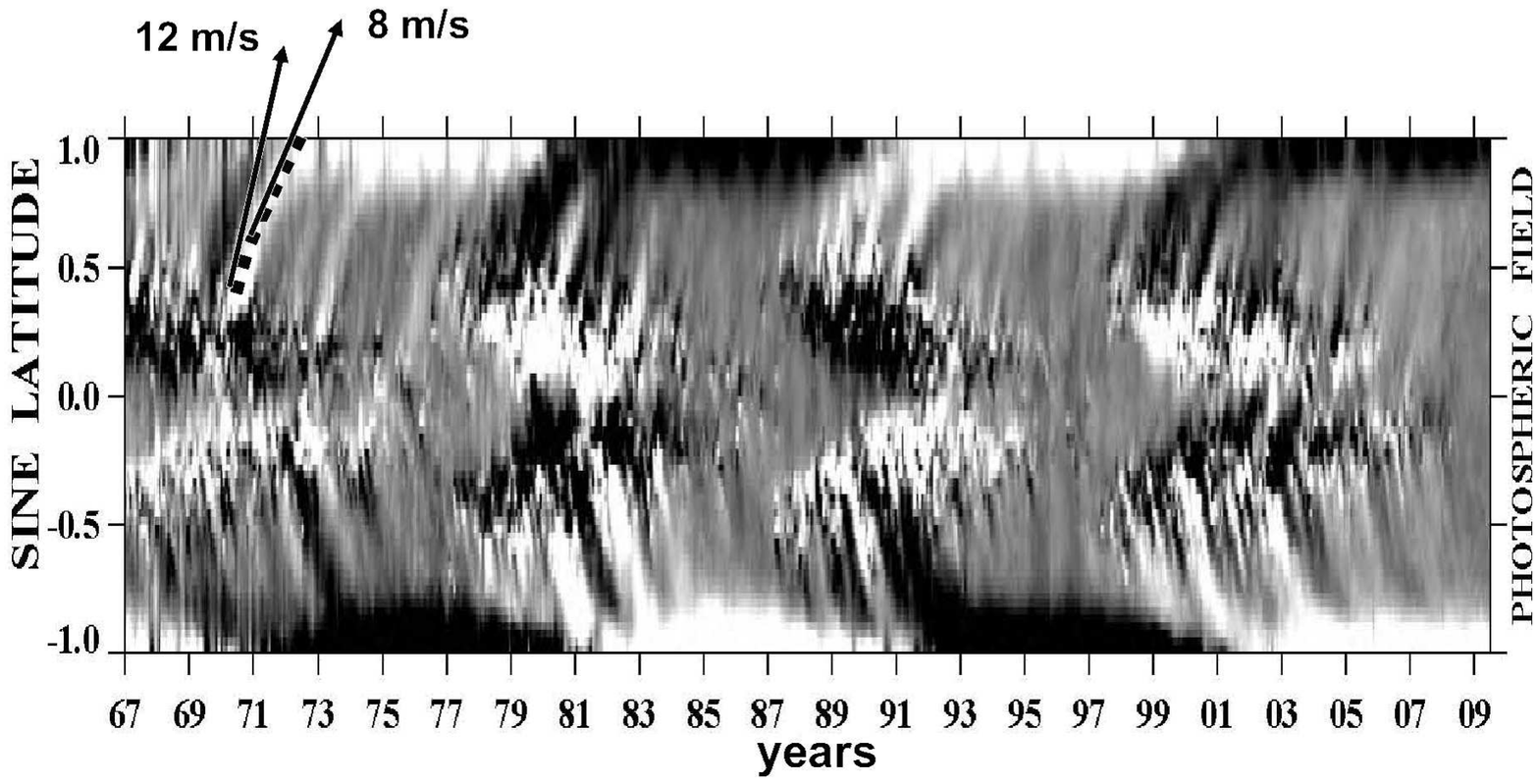}}
\caption{{\itshape Longitudinally averaged photospheric field observed at the Mount Wilson Observatory since 1967, showing surges of flux migrating poleward from the sunspot belts during each sunspot cycle.  The dotted curve marks a positive-polarity surge whose slope (converted from sine latitude to latitude) changed from 12 m s$^{-1}$ to 8 m s$^{-1}$ as it moved northward during 1970--1972.}}
 \end{figure}

Now we know that diffusion and flow both contribute to the transport \citep{DSB_84,WNS_89} and that both terms must be considered when interpreting the slopes of the surges \citep{WRS_09}.  As described by \cite{SWD_89}, supergranular diffusion provides an effective flow that is proportional to the surface gradient of magnetic flux density.  Thus, in Figure~10, the initial slope of the surge marked by a dotted line is 12 m s$^{-1}$, but diffusion may contribute $\sim$30$\%$ of this value, depending on the magnitude of the flux gradients in the sunspot belts.  At higher latitudes the gradients are smaller and the 8 m s$^{-1}$ slope may more closely represent the true speed of meridional flow.

The flow speed can be inferred through its influence on the magnetic field distribution.  At low latitudes, the competition between poleward flow and equatorward diffusion from the sunspot belts determines the amount of leading-polarity flux that reaches the equator and is annihilated by its counterpart in the other hemisphere. Consequently, this competition also determines the amount of unbalanced trailing-polarity flux that remains in each hemisphere for reversing the polar field.  At high latitudes, the competition between poleward flow from the sunspot belts and equatorward diffusion from the polar region determines the latitudinal shape of the polar field.  The observed topknots of flux \citep{SDS_78} are consistent with a flow profile that decreases linearly or quadratically toward the poles \citep{DSB_84,SWD_89}.

A new era of understanding occurred when \cite{WLS_02} found that a small variation of flow speed would be sufficient to regulate the polar-field reversal if the speed were correlated with the strength of the sunspot cycle.  In a strong cycle, a faster flow would produce less unbalanced trailing-polarity flux and cause the polar fields to fluctuate until the unbalanced flux arrived there late in the cycle.  In a weak cycle, a slower flow would compensate by producing more unbalanced trailing-polarity flux.

Recent simulations suggest that a small ($\sim$$15\%$) increase of flow speed may be responsible for the currently weakened polar and interplanetary fields
\citep{SL_08,WRS_09}.  On the other hand, the increased depth and length of this minimum probably result from changes in the subsurface return flow, as suggested by the reduced migration speed of the torsional oscillations prior to the start of cycle 24 \citep{HOW_09}. 

\section{It's your grandfather's solar minimum!}

Having examined the present solar minimum and compared it with previous minima, we find differences which suggest that `It's not your father's solar minimum' (to paraphrase the Oldsmobile advertisement).  However, this minimum does contain similarities to older minima during the past 100 years or more.  Perhaps it's your grandfather's solar minimum.

%\begin{quote}
%\verb"\begin{figure}"\\
%\verb"\plotone{"\arg{eq2065b.pdf}\verb"}"\\
%\verb"\caption{"\arg{Photospheric magnetic field}\verb"}"\\
%\verb"\end{figure}"
%\end{quote}

%%% Top level section head (remove "%" symbol)
%\subsection{}   %%% Second level section head (remove "%" symbol)
%\subsubsection{}   %%% Lowest level section head (remove "%" symbol)
%\section*{}    %%% Unnumbered top level section head (remove "%" symbol)
%\subsection*{}   %%% Unnumbered second level section head (remove "%" symbol)

\acknowledgements %%% Text of acknowledgements runs on after this command.
I am grateful to Y.-M. Wang (NRL) for several useful discussions and for providing
material for Figures~6, 8, and 10.  Sunspot numbers in Figure~3 came from
www.ngdc.noaa.gov/stp/SOLAR/ftpsunspotnumber.html. I thank David Hathaway (NASA/MSFC)
for permission to use his butterfly diagram (solarscience.msfc.nasa.gov). I thank
Rachel Howe (NSO) for providing material for Figure~5 and for several useful
discussions.  Frank Hill, Rudi Komm, and Irene Gonz{\'a}lez-Hern{\'a}ndez (all at NSO)
provided helpful comments about helioseismology, and Roger Ulrich (UCLA) provided MWO
measurements of torsional oscillations and episodic poleward surges which I continue to
appreciate.  Financial support was provided by NASA and NRL.
%%% THE BIBLIOGRAPHY
%%%
%%% CONSULT SECTION 3 OF "INSTRUCTIONS FOR AUTHORS" FOR HOW TO USE NATBIB.
%%% AUTHORS ARE ENCOURAGED TO USE EITHER THE "THEBIBLIOGRAPY" ENVIRONMENT
%%% BY UNCOMMENTING (DELETING THE "%" SYMBOL) THE COMMANDS BELOW, OR BY
%%% USING THE BIBTEX ENVIRONMENT. TO FIND OUT WHICH IS APPLICABLE TO YOUR
%%% CONTRIBUTION, CONSULT THE VOLUME EDITORS FOR YOUR PROCEEDINGS.
%%%

%\bibliography{refs}
%\bibliographystyle{apj}

\end{document}